**Matters arising**

from Duer, M. et al., Observation of a correlated free four-neutron system, Nature 606, 678 (2022)

## New results on the tetraneutron, seen in context.

Thomas Faestermann

Technische Universität München, James-Franck-Str. 1, 85748 Garching, Germany
Email: thomas.faestermann@ph.tum.de

In a recent publication[1] an experiment is described which searched for a resonance in the four-neutron system. In a convincing way the knockout of an α-particle off $^8$He nuclei by protons has been measured and the missing mass of the remaining four neutrons has been calculated from the measured four-momenta of α-particles and protons. The missing mass spectrum shows a bump of events corresponding to excitation of the four-neutron system to the continuum. In addition, there is a peak near zero missing mass that is convincingly interpreted as a resonance with excitation energy of the four neutrons E*= 2.37(58) MeV and a width of a Breit-Wigner distribution of Γ=1.75(37) MeV [A].

Unfortunately, in the Duer et al. publication[1] no reference has been made to our paper[2] which had appeared six months earlier, but after the submission of their manuscript to Nature. At least, the corresponding News and Views article[3] should have mentioned our measurement[2], since that[3] was presumably written after acceptance of the Duer et al. paper[1].

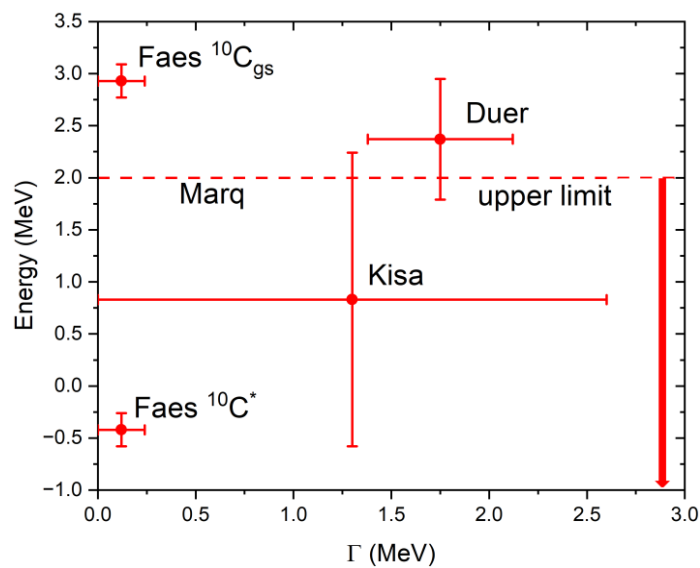

Fig. 1. Results of the four experiments [1,2,4+5,6] with respect to energy, available for the decay into four neutrons, and width Γ of a claimed tetraneutron state. All, except Duer et al.[1], quote only an upper limit for the width. The value favored by us[2] is the bound state at negative energy corresponding to the ejectile $^{10}$C* in the first excited state and a bound tetraneutron, because a width smaller than 0.24 MeV of a state unbound by 2.93 MeV would be unreasonably small.

In Fig. 1 I show the results of the four experiments[1,2,4+5,6] with respect to excitation energy and width which claimed in this century the observation of a bound or nearly bound four-neutron system. Marqués et al.[4], using fragmentation of $^{14}$Be nuclei, had claimed a bound tetraneutron but had to admit later on[5], that their observation could also be explained by an unbound tetraneutron with the decay neutrons emitted in a narrow cone and then just could quote an upper limit for the decay energy. Kisamori et al.[6] have used a double charge-exchange reaction $^4$He($^8$He,$^8$Be) and find a peak in the missing mass spectrum with an upper limit for the width and an excitation energy which also allows negative values, thus allowing also a bound tetraneutron. In our three-proton pickup reaction $^7$Li($^7$Li,$^{10}$C) we find a peak corresponding to an excitation E*=2.93 MeV for the $^{10}$C+4n system. This energy would be compatible with the Duer et al. value[1], but for the width $\Gamma$ we get an upper limit of 0.24 MeV, compatible with zero width; therefore, a disagreement with the Duer et al. observation[1]. Since a width $\Gamma$<0.24 MeV for a 2.93 MeV resonance is unrealistic we attributed[2] our peak to an excited $^{10}$C* ejectile and a tetraneutron bound by 0.42(16) MeV. If both experiments are correct, they cannot observe the same state of the tetraneutron. Looking at the excitation energies of the first excited states of the heavier even-even N=4 nuclei $^6$He, $^8$Be, $^{10}$C and $^{12}$O, formed by the coupling of the two $0p_{3/2}$ neutrons to $2^+$, I find an average excitation energy of 2.54 MeV. Therefore, the energy difference of the Duer et al.[1] result and ours[2] of 2.79(60) MeV would be compatible with the situation that Duer et al. were observing the first excited state of the tetraneutron and we the ground state. A small population of the ground state might be hidden under their background. And in our experiment population of the first excited state might be unobservable due to its large width. Still, more theoretical and experimental efforts seem necessary to clarify the properties of the tetraneutron.

---

[A] The value of Γ=1.75 MeV quoted by Duer et al.[1] gives rise to some doubt. With a ruler I get from their Fig. 3 as full width at half maximum (FWHM) a value of 5.5 MeV. The convolution of a Lorentzian (width Γ=FWHM) with a Gaussian (width σ) gives a Voigt profile. An approximation for the width of a Voigt profile is $FWHM_{Voigt} = \Gamma/2 + \sqrt{(\Gamma/2)^2 + 8ln(2)\sigma^2}$. From a value for σ=1 MeV[1,7] and a FWHM$_{Voigt}$=5.5 MeV results Γ=4.5 MeV!